# Waveform-Selective Mantle Cloaks for Intelligent Antennas

Stefano Vellucci, *Student Member, IEEE*, Alessio Monti, *Senior Member, IEEE*, Mirko Barbuto, *Senior Member, IEEE*, Alessandro Toscano, *Senior Member, IEEE* and Filiberto Bilotti, *Fellow, IEEE*

*Abstract*—We present the design of an innovative wire antenna able to automatically hide or reveal its presence depending on the waveform of the received/transmitted signal. This unconventional behaviour is achieved through the use of a novel waveform-selective cloaking metasurface exploiting a meander-like unit cell loaded with a lumped-element circuit capable to engineer the scattering of the antenna depending on the waveform of the impinging signal. Due to the time-domain response of the lumped-element circuit, the antenna is able switching its scattering behaviour when interacts with either a pulsed wave (PW) or a continuous wave (CW) signal. The proposed configuration paves the way to a new generation of cloaking devices for intelligent antenna systems, extending the concept of antenna as a device capable to sense the external environment and change its electromagnetic behaviour accordingly.

*Index Terms*—Electromagnetic cloaking, mantle cloaking, metasurfaces, reconfigurable metasurfaces, waveform selective metasurfaces, receiving antennas, scattering, dipole antennas.

## I. Introduction

IN the last decades, metasurfaces, the two-dimensional counterpart of metamaterials, have drawn great attention thanks to their ability to manipulate the electromagnetic field by introducing a field discontinuity across an interface. Metasurfaces are usually composed of passive unit cells, made of metallic or dielectric inclusions small compared to the operation wavelength, and are characterized by conformal designs and lower losses compared to metamaterial, especially at RF and microwave frequencies. Thanks to these advantages, different research groups have put great efforts in designing novel and intriguing devices based on metasurfaces including beam shapers [1],[2], filtering inclusions [3]-[5], invisibility cloaks [6],[7], scattering manipulating devices [8],[9] and many other functional components [10]-[12].

In particular, in the frame of electromagnetic cloaking, A. Alù has proposed the use of thin metasurfaces to hide a coated target object [6]. Exploiting the field scattered by the coating metasurface, the field scattered by the object can be compensated, leading to a dramatic reduction of the scattering signature and, thus, to a cloaking effect. When using this technique, referred to as *mantle cloaking*, the cover surrounding the object does not inhibit the interaction between the external field and the object itself. For this reason, and for the straightforward design of the coating cloaking metasurface [13]-[16], the interest in this technique by antenna engineers has increased, leading to new applications in both terrestrial and satellite scenarios [17]–[24]. In this context, one of the most investigated application of mantle cloaking relies on the use of properly engineered conformal metasurfaces to reduce the blockage effect and the mutual coupling between electrically close antennas. This approach has been introduced in [17] and validated experimentally in [20],[21] for the case of two monopole antennas and, then, extended to other antennas [23].

Moreover, the concept of power-dependent mantle cloak for antenna arrays has been recently introduced [24]. It has been shown that the cloaking behavior of a capacitive metasurface can be switched on/off by loading the gaps between the metallic strips of the metasurface and exploiting the non-linearity characteristic of diode pairs. This advanced functionality of the metasurface has enabled the design of new power-dependent antennas able to modify their radiative characteristics depending on the input power level. Electronic elements can be also used to broaden the frequency bandwidth of mantle cloaking devices. For instance, in [25], it has been shown that broadband scattering suppression can be achieved by exploiting active negative impedance converters loading a cloaking metasurface.

The use of electronic elements to load metasurfaces for achieving advanced electromagnetic functionalities is not limited to cloaking, but, rather, is a common trend of the current research in electromagnetics, making metasurfaces tunable and reconfigurable, to enable the design and realization of real-time controlled devices changing dynamically their electromagnetic properties. In particular, electrically tunable metasurfaces have been successfully demonstrated exploiting semiconductors, liquid crystals, graphene, and more exotic solutions based on liquid metals [26]-[33], for the implementation of different practical

Manuscript received April XX, 2019; accepted XXX XX, 201X. Date of publication XXXX XX, 201X; date of current version XXXX XX, 201X. This work has been developed in the frame of the activities of the research contract MANTLES, funded by the Italian Ministry of Education, University and Research as a PRIN 2017 project (protocol number 2017BHFZKH). (*Corresponding author: S. Vellucci*)

S. Vellucci, A. Toscano, and F. Bilotti are with the Department of Engineering, ROMA TRE University, 00146 Rome, Italy (e-mail: stefano.vellucci@uniroma3.it).

A. Monti and M. Barbuto are with the Niccolò Cusano University, 00166, Rome, Italy.

Color versions of one or more of the figures in this communication are available online at http://ieeexplore.ieee.org.

Digital Object Identifier X

applications, such as tunable reconfigurable antennas [34], broadband metasurfaces for transmission and reflection control [35],[36], tunable meta-lenses [37], and tunable perfect absorber/reflector [38].

Among the different unconventional effects enabled by metasurfaces loaded by circuit elements, recently, the group of. D. Sievenpiper has introduced the concept of waveform-dependent (or -selective) metasurfaces [39]-[42]. Introduced for designing metasurface-based absorbers capable to absorb high-power pulse energy while allowing propagation of weak signals [39], waveform-selective metasurfaces have been then exploited for designing devices exhibiting different transmitting/reflective properties, depending on the waveform of the impinging signals [40],[41], i.e. able to distinguish between a short pulsed wave (PW) or a continuous wave (CW) signal. To induce this unconventional waveform-selective effect capacitor-based or inductor-based lumped elements circuit are exploited. Moreover, the possibility to exploit more complex circuit based on active elements, such as jFET [42], provides additional design flexibility in electromagnetic problems, enabling new kinds of microwave applications.

Inspired by these works, the aim of this paper is to further expand this emerging field of reconfigurable/tunable devices by introducing the design of an intelligent antenna able to hide or reveal its presence, automatically changing its electromagnetic visibility state, depending on the received/transmitted signal waveform. In particular, we show how a properly designed mantle cloaking device is able to change the in band-scattering signature of a wired antenna depending on the waveform of the received/transmitted signal, for designing an antenna that operates correctly when excited by continuous waves (CW), while exhibiting minimum scattering when illuminated by a short pulse (PW).

The paper is organized as follows. In Section II, we introduce our idea of an antenna able to sense the external electromagnetic environment and change automatically its visibility state. The potential impact of this solution is discussed, and a possible application is suggested. In Section III, we report the design procedure of the lumped-element circuit exploited to induce this unconventional cloaking behavior. In Section IV, the final design of the waveform-selective cloaking metasurface is presented and its performance evaluated varying the waveform of the impinging signal and showing its different cloaking properties. Finally, the performance of the antenna coated by the cloaking device is presented and discussed.

## II. INTELLIGENT INVISIBLE ANTENNA CONCEPT

The intelligent invisible antenna we propose here is an antenna able to hide/show itself depending on the waveform of the received signal. In particular, the antenna is coated with a mantle cloak able to suppress the scattering signature of the antenna at its resonant frequency $f_0$ when interacting with a pulse wave (PW), while the cloaking effect is automatically switched-off when the antenna is illuminated by a continuous wave (CW).

As is well known, a metasurface can be implemented at microwave frequencies through a patterned metallic surface with a subwavelength period. This structure can be modeled using a homogenized physical quantity known as the average surface impedance $Z_s$ ($\Omega$/sq) [43]-[45]. Thus, to achieve the described unconventional waveform-dependent cloaking behavior, we propose to design a cloaking metasurface exhibiting a surface impedance that depends on the waveform of the impinging signal. Since the mantle cloaking theory allows suppressing the scattering signature of a wire antenna by coating it with a properly engineered metasurface [6],[17]-[24], the metasurface to design here should be able to match the cloaking condition ($Z_s = Z_s^{cloak}$) for a PW excitation, while exhibiting a different $Z_s$ when the antenna is illuminated by a CW. This unusual cloaking behavior is achieved by loading the metasurface with a waveform-dependent circuit exploiting passive lumped-elements. Considering that the circuit does not require any active element, the antenna coated by this waveform-dependent mantle cloak is able to adapt its visibility state passively and automatically.

To better clarifying this aspect, a sketch of the working principle of such a waveform-selective mantle cloak when applied to a wired dipole antenna is illustrated in Fig. 1. When a PW illuminates the coated antenna, the cloaking behavior is turned on, making the antenna invisible. Differently, when a CW signal impinges onto the coated antenna, the cloaking effect is turned off, and the radiator works as a regular antenna. It is worth underling that the proposed waveform-dependent mantle cloaking device is designed to work for two signals operating at the same frequency $f_0$, i.e., for a PW characterized by a main component at the same modulating frequency of the CW signal. Thus, our work intrinsically differs from previous cloaking solutions exploiting Fano resonances to introduce an on/off cloaking behavior at different frequencies [46]-[48].

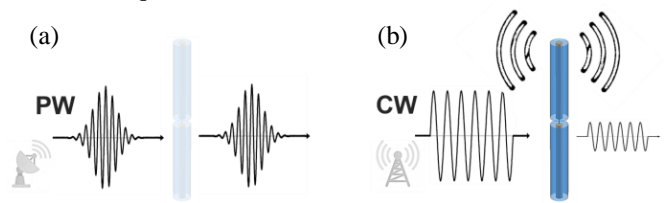

Fig. 1. Sketch of the waveform-dependent cloaking concept applied to a wire antenna. (a) The cloak switches on if illuminated by a pulse signal (PW), and the antenna becomes invisible, while (b) the cloak is in an off-state if illuminated by a continuous signal (CW), and the antenna operates in the usual way in both reception and transmission.

This waveform-dependent mantle cloak could be of particular interest for defense applications, in particular in the frame of microwave radars. In radar systems, in fact, the electromagnetic waves are usually radiated as short pulses (PW). The pulsed waveform is required to reach the target and allowing a portion of the energy coming back to the same antenna (or a different antenna if a bistatic system is used) before the next burst of waves is transmitted [50]. Thus, exploiting the proposed waveform-dependent cloak it is possible designing communication antennas able to talk with a base-station while becoming automatically invisible to the signal of a detecting radar. In this way, it would be possible to mitigate the problem of the strong visibility of antennas at their resonant frequency leading to a strong scattering signature [50]. Moreover, it is worth noticing that the pulse width of radar signals operating at microwaves is typically of the order of microseconds and, thus, the pulsed wave is characterized by a frequency bandwidth that is rather small compared to the typical operative bandwidth of mantle cloaks [6].

Finally, we stress that in previous works on the mantle cloaking technique for antenna applications [17]-[24], cloaking devices have been designed to reduce only the scattering of the antenna outside its operation frequency band. In fact, as known from the fundamental theory of passive cloaking [18], an antenna cannot at the same time be cloaked at its own resonant frequency and continue operating in transmission/reception. As shown Fig. 2, without loss of generality for the case of a resonant dipole wire antenna, an antenna cloaked at its own resonant frequency would be strongly mismatched to the load. Therefore, the antenna would be unable to efficiently receive or transmit.

From a physical point of view, this effect is consistent with the limitations coming from the optical theorem [51] and the energy conservation principle. In fact, the optical theorem reads [52]:

$$\sigma_{ext} = \sigma_{abs} + \sigma_{scat} = Im[s_\theta(0,0)]\frac{\lambda_0}{\pi} \quad (1)$$

where $\sigma_{ext}$ is the total extinction cross section of the object, $\sigma_{abs}$ the absorption cross section, $\sigma_{scat}$ the scattering cross section and the quantity $Im[s_\theta(0,0)] \cdot \lambda_0/\pi$ the normalized scattering amplitude parallel to the impinging filed in the forward direction $s_\theta(0,0)$. Eq. (1) means that an object with zero or weak total scattering (e.g., $\sigma_{scat} = 0$ and, thus, $s_\theta(0,0) = 0$) must exhibit zero or weak total absorption (e.g., $\sigma_{abs} = 0$) i.e., some absorption is always required in order to have nonzero scattering. Since the optical theorem applies to the scattering of any object illuminated by a plane wave, which is linearly polarized in the specific case of a wire antenna, it implies that an antenna characterized by zero scattering (e.g., cloaked at its own resonant frequency) is unable to receive/transmit since it would be characterized by zero absorption. This statement is also consistent with the limit of absorption from a receiving antenna, discussed in [53].

However, the proposed waveform-selective cloak allows turning this fundamental limitation around, since enables the design of an antenna invisible to a radar emitting PW signals, while keeping its proper behavior for the CW signals, needed for communication purposes at the same frequency of operation.

### III. DESIGN OF THE WAVEFORM-SELECTIVE CIRCUIT

To design the waveform-dependent mantle cloak, we first need to design the waveform-selective circuit. Following the approach discussed in [39]-[42], we have designed the lumped-element circuit shown in Fig. 3, consisting of a 4 diodes full-bridge rectifier and a series combination of an inductance (L) and a resistor (R) connected in parallel to the input ports (P1 and P2).

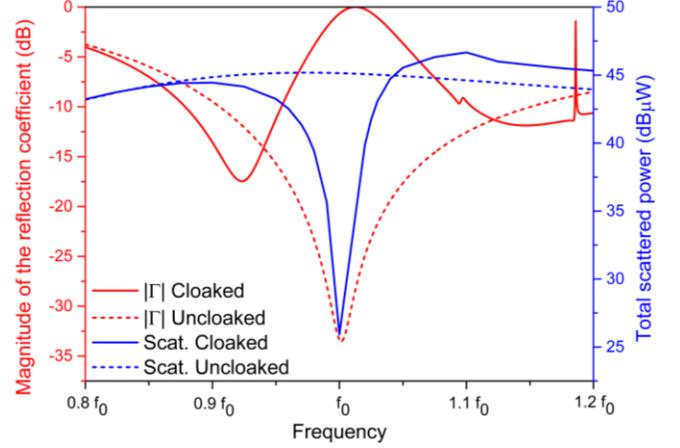

Fig. 2. Example of the total scattered power and of the reflection coefficient of a dipole antenna in the uncloaked and cloaked scenarios (the cloak is designed to work at the resonant frequency $f_0$ of the antenna). The dipole is strongly mismatched in the cloaked scenario, while it is well matched in the uncloaked case.

Due to the engineered time-domain response of the lumped-elements circuit, the points P1 and P2 can be switched from an open circuit (o.c.) to a short circuit (s.c.) condition (and viceversa), depending on the waveform of the signal and, in particular, on its pulse width ($\Delta t$). When an oscillating signal is applied between the two ports, the fundamental frequency of the signal is converted into an infinite set of frequency components by the full-bridge rectifier, with the main one at zero frequency. Therefore, a main DC component appears. Due to the presence of the inductance L, initially the flow of the induced current is opposed, and the two ports P1 and P2 of the circuit are in o.c. However, if the signal is characterized by a $\Delta t$ long enough compared to the time constant of the circuit ($\Delta \tau = L/R$), the opposing effect of the inductance gradually weakens, and current starts flowing between the two ports through the resistance R.

It is worth noticing that the presence of the rectification process due to the diodes is essential for the correct operation of the circuit. The non-linear characteristics of the diodes allow exploiting the transient time-domain response of the inductance L generating an opposing electromotive force that weakens after some time (Fig. 3 (b)). Without the DC conversion, the circuit would behave in a similar way for both the PW and CW signals.

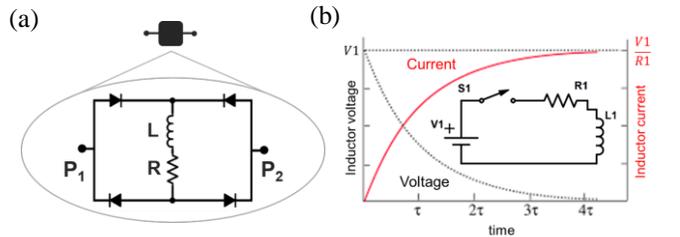

Fig. 3. (a) Waveform-selective circuit consisting of a four diodes rectifier and a series combination of an inductance (L) and a resistor (R) connected in parallel to the input ports (P1 and P2). (b) Typical transient response of an inductor.

To confirm the expected behavior of the waveform-selective circuit, numerical simulations have been carried out by using the commercial software CST Studio Suite™ [54], and the results are shown in Fig. 4. The performance of the circuit has been analyzed in terms of the currents flowing between P1 and P2. In particular, current probes have been placed at the input ports and two different signal sources have been used to excite the circuit. As shown in Fig. 4 (a), a cosine function at 3 GHz has been used as CW source. Differently, a broadband Gaussian pulse having a pulse width $\Delta t = 0.7$ μs modulating a cosine at 3 GHz has been used as PW source. This signal has been chosen because it can be assumed to be closer in terms of the pulse width and frequency spectrum to a realistic radar threat for antennas operating at microwave frequencies.

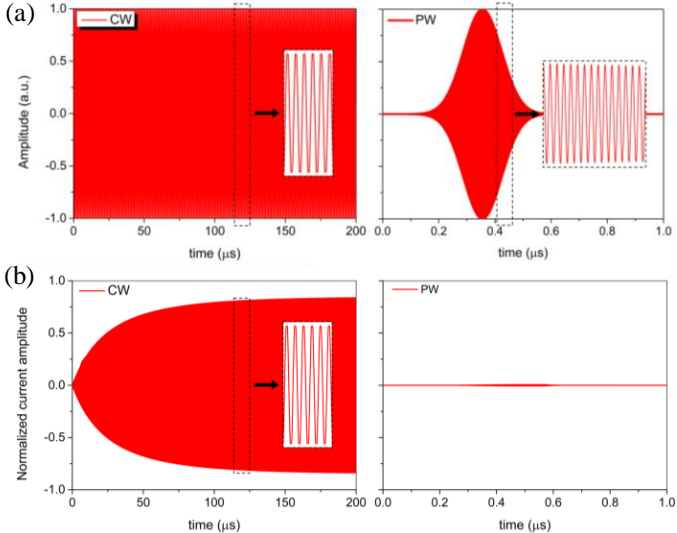

Fig. 4. (a) Exciting signal waveforms. The CW is a cosine function at 3 GHz, while the PW is a Gaussian pulse at 3 GHz characterized by a $\Delta t = 0.7$ μs. (b) Time domain response of the circuit given in terms of the currents measured at P1 and P2 in Fig.3 for a CW signal and a PW signal. In the insets, details of the waveforms.

In Fig. 4 (b)-(c) the time-domain response of the current amplitudes flowing between P1 and P2 in the cases of CW and PW excitations are shown. Please, observe that the current amplitudes have been normalized to the current peak value of the CW case. As can be appreciated, in the case of CW (Fig. 4 (b)), after an initial rise time, a steady state of conduction between the two ports is achieved and a strong current is induced. Thus, ports P1 and P2 are in s.c. condition. On the contrary, when the PW is excited (Fig. 4 (c)) weak currents (around 3 orders of magnitude less compared to the previous case) flow through the circuit, indicating the o.c. condition of the two ports.

Please, note that in the CW case the simulation has been terminated after 200 μs to reduce the simulation time and the computational cost. However, as it can be seen in Fig. 4 (b), the current curve nearly saturated at 200 μs, indicating that the steady state condition has been met. It is also worth mentioning that the values of the lumped elements in the circuits have been chosen to maximize the differences between the two cases and have the following values: L = 800 μH, R = 1 Ω. In particular, the value of the resistor R has been chosen to minimize losses and to be close to the typical value of the internal resistance of inductor chips. Ideal diodes have been instead used to speed up the simulation time. However, in all the performed simulations and in the co-simulations described below, non-linear effects and generation of higher-order harmonics have been fully taken into account.

Finally, we also remark that an opposite time-domain response can be achieved exploiting a complementary RC circuit. In particular, as proved in [42], by replacing the series combination between the inductance and the resistor with a parallel combination of a capacitance and a resistor, it is possible to design a circuit able to make a short circuit between P1 and P2 for a PW and an open circuit for a CW. This would grant the possibility to design a waveform-dependent mantle cloak able to suppress the scattering signature of an antenna when interacting with a CW, while restoring the visibility level of the antenna when illuminated by a PW, for achieving a reciprocal waveform-dependent cloaking effect compared to the one proposed here.

## IV. DESIGN OF THE WAVEFORM-DEPENDENT MANTLE CLOAK

### A. Meander-Like Cloaking Metasurface

To design the intelligent invisible antenna, we need to engineer a waveform-dependent mantle cloak able to suppress the scattering signature of the antenna. In particular, the first problem we deal with is the reduction of the radar signature of a dipole antenna operating at $f_0 = 3$ GHz.

Following the theory of mantle cloaking, the scattering signature of a wire antenna can be suppressed by coating it with a properly designed metasurface characterized by a specific cloaking value of the surface impedance (i.e., $Z_s = Z_s^{cloak}$). However, to achieve the waveform-dependent cloaking behavior, the value of $Z_s$ should vary between the cloaking value ($Z_s^{cloak}$) and a value able to restore the natural radiative and electrical characteristics of the antenna.

Since we would like to suppress the scattering signature of a dipole at its own resonant frequency $f_0$, an inductive metasurface is required to cloak the antenna [14],[16]. To achieve this goal, we have slightly modified a classical vertical metallic strip metasurface by using a meander-like inductive metasurface unit cell, as the one depicted in Fig.5 (a)-(b). Here, the value of the surface impedance depends on the parameter $w$, as for standard vertical inductive metallic strips [16], but also depends strongly on the parameter $k_2$. A variation in the value of $k_2$ results in a significant change in the length of the electrical path of the induced currents. Thus, by short-circuiting the points P1, P2, where the waveform-selective circuit will be placed, it is possible to change the metasurface $Z_s$ and transform the inductive meander-like unit cell in an equivalent conventional inductive vertical strip (Fig. 5 (c)), whose value of the $Z_s$ depends just on $w$.

The final design of the cloaking metasurface is shown in Fig. 5 (b). In particular, the dipole antenna resonating at 3 GHz has a radius $a = \lambda_0/50$, a length $L = \lambda_0/39$, and a gap between the two branches $g = \lambda_0/100$. The cloaking metasurface, instead, is made of 3 meander-like unit cells, placed on a dielectric hollow cylinder with permittivity $\varepsilon_r = 6$ and radius $a_c = 1.3 \cdot a$. The electrical dimensions of the meander-like unit cells are: $w = \lambda_0/77$, $h = \lambda_0//19$, $k_1 = \lambda_0//50$, $k_2 = \lambda_0//20$ and $D = \lambda_0//18.4$ (see Table I for a summary of the geometrical dimensions).

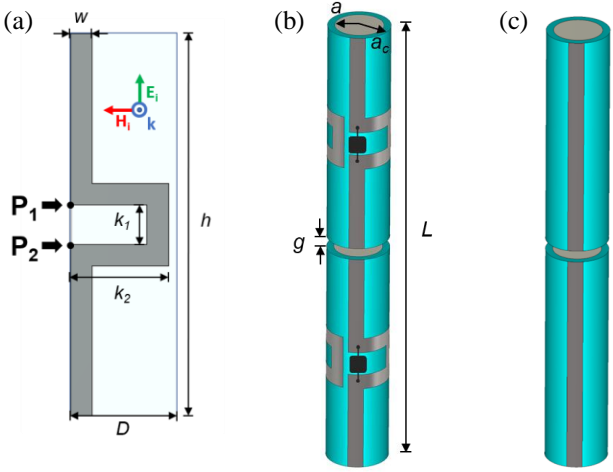

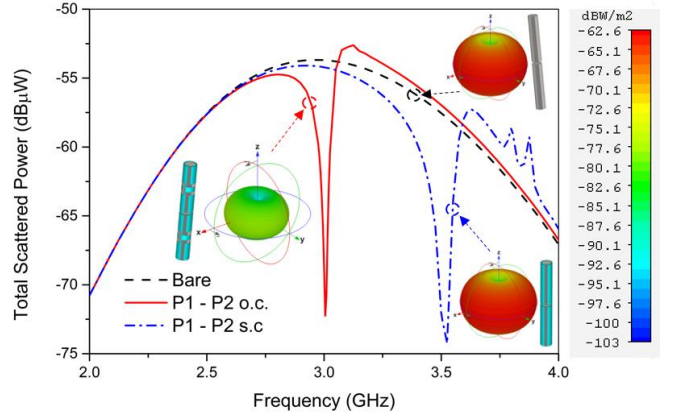

Fig. 5. (a) Unit cell of the inductive meander-like metasurface excited by a plane wave whose electric field is parallel to the strip axis. The picks P1 and P2 are the points where the loading circuits its applied. (b) Dipole antenna coated by the waveform-dependent mantle cloak loaded with the lumped-element circuits. (c) Equivalent geometry of the mantle cloak short-circuiting the points P1, P2. The electrical dimensions of the structures are reported in Table I.

The cloaking behavior of the designed metasurface in the circuit unloaded case is confirmed by the full-wave numerical results in Fig. 6. As can be appreciated, when P1, P2 are in o.c. condition and $Z_s = Z_s^{cloak} = j29.9$ $\Omega$/sq, the metasurface behaves as a cloaking device suppressing the scattering signature of the antenna at its own resonant frequency (Fig. 6, red continuous line). Differently, when P1, P2 are in s.c. and the surface impedance changes to $Z_s = jX_s = j20.1$ $\Omega$/sq, and the cloaking peak is frequency shifted, restoring the scattering performance of the antenna at $f_0$ (Fig. 6, blue dash-dotted line). In this case, the scattered power returns to the level when the coating cloak is not present (Fig. 6, black dashed line).

TABLE I

ELECTRICAL DIMENSIONS OF THE MEANDER-LIKE METASURFACE UNIT CELL AND OF THE DIPOLE ANTENNA

| Symbol | Quantity | Value |
| --- | --- | --- |
| $w$ | Width of the strip | $\lambda_0 / 77$ |
| $h$ | Height of the strip | $\lambda_0 / 19$ |
| $k_1$ | Height of the meander | $\lambda_0 / 50$ |
| $k_2$ | Depth of the meander | $\lambda_0 / 20$ |
| $D$ | Unit cell periodicity | $\lambda_0 / 18.4$ |
| $L$ | Length of the dipole | $\lambda_0 / 39$ |
| $a$ | Radius of the dipole | $\lambda_0 / 50$ |
| $g$ | Gap of the dipole | $\lambda_0 / 100$ |
| $a_c$ | Radius of the cloak | $1.3 \cdot a$ |

Fig. 6. Comparison between the performances of the dipole antenna in terms of the total scattered power in the scenarios without the cloak (bare case – black dashed line), with P1, P2 in o.c. condition (cloaked case – red continuous line) and with P1, P2 in s.c. condition (uncloaked case – blue dot dashed line). In the inserts, the scattered patterns of the antenna at $f_0$ and the equivalent geometries in the three cases are shown.

In Fig. 6 is also possible to appreciate the 3D scattering patterns of the antenna in the three cases evaluated at $f_0$, confirming the strong scattering reduction of the antenna when P1, P2 are in o.c. and the unaltered scattering characteristic when P1, P2 are in s.c. condition. It is also worth mentioning that, as expected, the scattering signature of the antenna is maximum around the resonant frequency.

*B. Waveform-Dependent Mantle Cloak and Antenna Performance*

To confirm the waveform-dependent cloaking behavior of the meander-like metasurface once loaded by the lumped-element circuit, a full-wave and circuit co-simulation analysis has been carried out, and the performance of the metasurface unit cell has been evaluated.

The simulation setup is reported in Fig. 7 (a) where the metasurface unit cell is loaded with a lumped port connected between points P1 and P2 excited by a normal incidence plane wave. Thanks to the co-simulation capabilities, the lumped port is able to fully represent the characteristics of the designed waveform-selective circuit in Fig. 3 (a), as if the circuit were directly placed between P1 and P2. This co-simulation setup allows drastically reducing the simulation time and properly evaluating the performance of the metasurface in presence of time-varying signals.

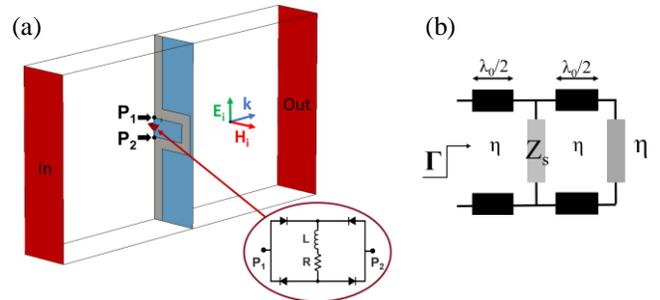

Fig. 7. (a) Geometrical sketch of the retrieval setup used for the evaluation of the metasurface $Z_s$ once loaded with the lumped-elements circuit. (b) Equivalent transmission line model.

Once the boundary condition for the unit-cell simulation are set, the structure can be conveniently represented by the equivalent transmission line model shown in Fig. 7 (b). In this simple model, the metasurface is represented by a shunted lumped element $Z_s$, while the spacing between the metasurface and the object is represented by a transmission line segment with length $\lambda_0/2$ and characteristic impedance $\eta$. The transmission line is also terminated to a loading impedance $\eta$. It is worth noticing that we have included two half-wavelength transmission line segments before and after the metasurface to neglect the contribution of possible evanescent fields near the interface.

The reflection coefficient $\Gamma$ of the equivalent transmission line model can be expressed as:

$$\Gamma = \frac{\eta - (Z_s||\eta)}{\eta + (Z_s||\eta)} \quad (2)$$

Thus, the numerical evaluation of the reflection coefficient from the co-simulation results allows retrieving the corresponding value of the metasurface surface impedance at $f_0$ when loaded by the circuit. In particular, in Fig. 8, it is shown the retrieved surface impedance when varying the pulse width $\Delta t$ of the exciting signal, compared to the surface impedance values when P1, P2 are in o.c. and s.c. conditions. In this case, a Gaussian pulse characterized by a pulse width $\Delta t$ modulating a cosine function has been considered.

As can be appreciated, when the pulse width of the signal is small compared to the time constant $\Delta\tau = L/R$ of the circuit (i.e., for a PW), the value of the surface impedance approaches the one obtained when P1, P2 are in o.c., (i.e., the cloaking value $Z_s^{cloak} = jX_s = j29.9$ $\Omega$/sq). Differently, when the pulse width of the signal increases and becomes larger compared to the time constant $\Delta\tau$ (i.e., for a CW), thanks to the time-response of the inductance in the circuit the value of the surface impedance approaches the one obtained by simply short circuiting P1, P2, (i.e., the value $Z_s = jX_s = j20.1$ $\Omega$/sq).

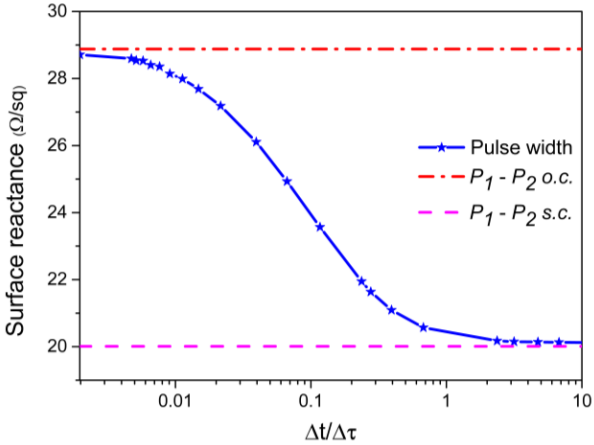

Fig. 8. Response at $f_0$ of the metasurface surface reactance ($X_s$) to a Gaussian pulse modulated by a cosine function at $f_0$ varying its pulse width $\Delta t$, in the cases of the metasurface loaded by the lumped-element circuit (blue continuous line), P1, P2 in o.c. condition (red dash dotted line), and P1, P2 in s.c. condition (purple dashed line).

Due to limitations and computational cost of the co-simulation method in evaluating the scattering performance of the antenna, to evaluate the final performance of the intelligent antenna, once coated by the designed waveform-dependent mantle cloak, the reciprocity principle has been exploited. Specifically, the magnitudes of the reflection coefficient of the coated antenna in case of a PW or a CW excitation have been analyzed. This equivalent method to evaluate the mantle cloak performance can be exploited since an antenna cloaked at its resonant frequency is characterized by a high value of the reflection coefficient, as mentioned in Section II.

As a reference of the reflection coefficient behavior in Fig. 9 we report the following cases. Fig. 9 (a) refers to the case when the mantle cloak is not present, i.e., the antenna is matched at $f_0$. Fig.9 (b) represents the case when the antenna is cloaked being coated by the circuit unloaded mantle cloak with P1, P2 in o.c. condition (i.e. $Z_s^{cloak} = jX_s = j29.9$ $\Omega$/sq). As can be appreciated the dipole is mismatched at $f_0$. Instead, when P1 and P2 are in s.c. condition (i.e., $Z_s = jX_s = j20.1$ $\Omega$/sq) the matching characteristics are restored, as shown in Fig. 9 (c), and the reflection coefficient behaves as if the cloak were not present.

These behaviors are confirmed when the mantle cloak is loaded with the waveform-selective circuit and the antenna is excited by either a PW or a CW signal, as reported in Fig. 9 (d). In particular, when the antenna source is a PW Gaussian pulse with $\Delta t = 0.7$ $\mu$s modulating a cosine at $f_0 = 3$ GHz the dipole is strongly mismatched at $f_0$ (blue line). The plot of the reflection coefficient almost overlaps the one in Fig. 9 (b), thus the antenna is effectively cloaked at $f_0$. On the contrary, when a CW cosine at $f_0$ is used as the source signal, the matching characteristics of the antenna are restored (red line) and the plot almost overlaps the curve in Fig. 9 (c).

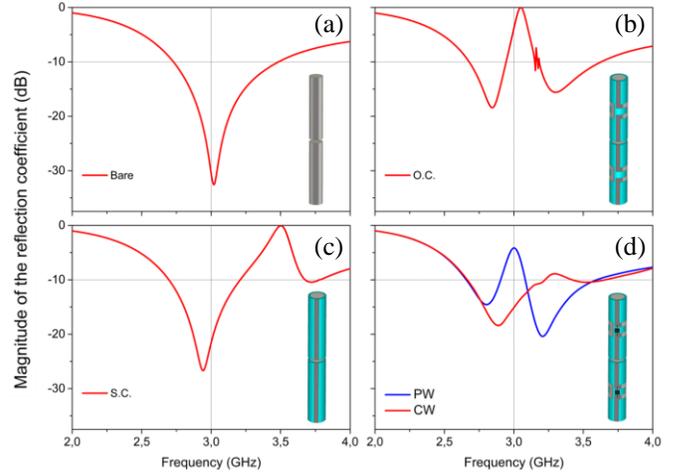

Fig. 9. Magnitude of the reflection coefficient of the dipole antenna in the scenarios (a) without the cloak (bare case), (b) with P1, P2 in o.c. condition (cloaked case), (c) with P1, P2 in s.c. condition (uncloaked case), (d) when excited by a PW (blue line) and a CW (red line) signal. In the insets, equivalent geometrical sketches of the antenna in the various conditions are shown.

It is worth noticing that in the PW and CW cases the reflection coefficient of the antenna has been computed by sampling its value at the same frequency of the modulated cosine when varying it from 2 to 4 GHz. This method was required to evaluate the reflection coefficient in the whole frequency range from 2 to 4 GHz, since both the PW and CW signals are narrowband and unable to cover the whole frequency spectrum of interest.

We stress that, due to full-wave simulation software limitations, we were unable to evaluate directly the cloaking performance of the whole antenna system when coated by the circuit-loaded mantle cloak. However, in the cloaked antenna case, the analysis of the reflection coefficient performance offers a reliable solution to these limitations.

Finally, as an additional proof of the waveform-selective cloaking effect, we report in Fig. 10, the 3D realized gain patterns of the coated antenna at $f_0$. In particular, in Fig. 10 (a) we show the realized gain pattern of the bare antenna without the cloaking metasurface, while in Fig. 10 (b)-(c) the realized gain patterns when the antenna is coated by the circuit-loaded mantle cloak and a CW or a PW is received/transmitted by the antenna are reported. Specifically, in Fig. 10 (b), the realized gain pattern of the coated antenna when excited by the CW cosine source is shown, while in Fig. 10 (c) the realized gain pattern of the antenna when excited by the PW Gaussian source is reported. As expected, the realized gain pattern in the CW case matches the one of the bare antenna, while in the PW case we observe a dramatic reduction of the realized gain due to the presence of a cloaking effect.

To conclude, we underline that, in order to experience the described waveform-dependent cloaking effect, the power received/transmitted by the antenna needs to be in a proper range. In particular, the incident power must be large enough to turn-on the diodes. This constrain can be relaxed reyling on the use of Schottky diodes. However, it is also worth noticing, that the pulsed wave signals used for radar application are usually characterized by a high-power level.

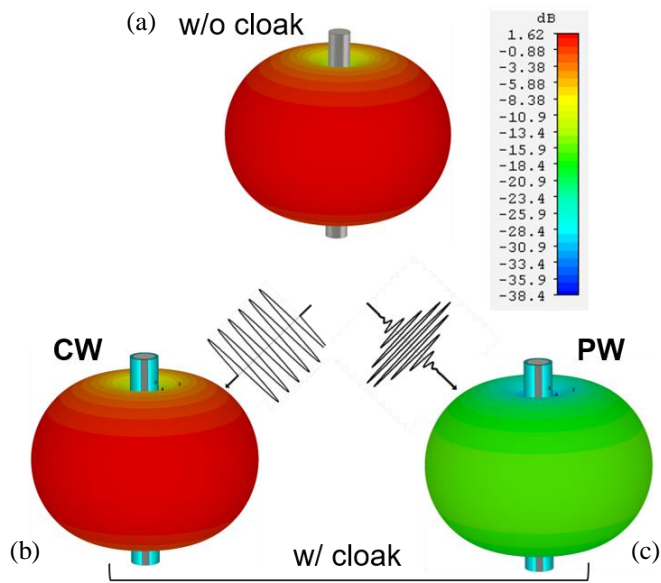

Fig. 10. Realized gain patterns of the antenna at $f_0$ in the scenarios (a) without the cloak (bare case), when the cloak is applied and the antenna is excited by (b) a CW or (c) a PW signal.

## V. CONCLUSIONS

In this work, we have presented the design of an antenna able to automatically hide/show itself depending on the waveform of the received/transmitted signal. The antenna exploits an unconventional waveform-dependent mantle cloak loaded with lumped-element circuits whose value of the surface impedance changes depending on the pulse width of the source signal. In particular, the waveform-dependent mantle cloak is characterized by a surface impedance able to cloak a dipole antenna at its own resonant frequency in presence of a PW. While the mantle cloak exhibits a different surface impedance when illuminated by a CW signal and the antenna operates correctly.

To achieve this unconventional cloaking behavior a lumped-element circuit able to behave as an open or short circuit according to the exciting signal waveform has been first designed. Then, a peculiar meander-like metasurface exhibiting different values of the surface impedance when loaded by the circuit in presence of a either PW or CW exciting signal has been proposed and its performance evaluated through a full-wave and circuit co-simulation. In particular, it has been proved that the surface impedance of the coating metasurface depends on the pulse width of the signal and switches from a value able to suppress the scattering signature of the coated antenna for a PW signal and a value able to restore the original characteristics of the antenna as if the cloak were not present.

The proposed configuration may pave the way to a new generation of intelligent and reconfigurable cloaking devices for antenna systems, expanding the growing field of metasurfaces by integrating functionalizing elements to enable advanced and unconventional operations.

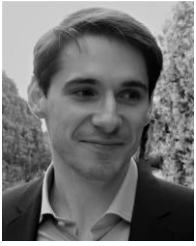

**Stefano Vellucci** (S'16) received the B.S. and M.S. (summa cum laude) degrees in electronic engineering from ROMA TRE University, Rome, Italy, in 2012 and 2015, respectively. In November 2015 he started his Ph.D. research program at the Department of Applied Electronic, Roma Tre University and graduated in April 2019.

Before starting his Ph.D., he was an Antenna Engineer with Elettronica S.p.A., Rome, Italy, where he designed, modeled and optimized antennas for military applications. His current research interests include the design and applications of artificially engineered materials and linear/reconfigurable metasurfaces for microwave components, with particular focus on the analysis and design of cloaking devices for antenna systems.

Dr. Vellucci is currently member of the *Virtual Institute for Artificial Electromagnetic Materials and Metamaterials* (METAMORPHOSE VI, the International Metamaterials Society), and of the *Institute of Electrical and Electronics Engineers* (IEEE). He has been serving as a Technical Reviewer of many high-level international journals related to electromagnetic field theory and metamaterial, and he was the recipient of the *IEEE AP-S Award of the Central-Southern Italy Chapter*, in 2019, of the *Outstanding Reviewers Award* from the IEEE Transactions on Antennas and Propagation, in 2018 and 2019, and of the *Leonardo-Finmeccanica Innovation Award for "Young Students"*.

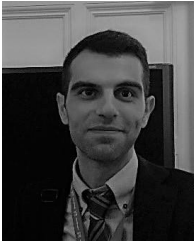

**Alessio Monti** (S'12–M'15–SM'19) was born in Rome, Italy. He received the B.S. degree (summa cum laude) and the M.S. degree (summa cum laude) in electronic and ICT engineering both from ROMA TRE University, Rome, in 2008 and 2010, respectively. From 2011 to 2013, he attended the doctoral school in biomedical electronics, electromagnetics and telecommunications engineering at ROMA TRE University. Currently, he is an assistant professor at Niccolò Cusano University, Rome, Italy, where he teaches courses on antenna and microwave theory.

Dr. Monti is member of the secretarial office of the International Association METAMORPHOSE VI, and of the Editorial Board of the journals *IEEE Transaction on Antennas and Propagation* and *EPJ Applied Metamaterials*. In 2019, he has been appointed as General Chair of the *International Congress on Artificial Materials for Novel Wave Phenomena – Metamaterials* and he has been serving as Chair of the Steering Committee of the same Congress series since 2017. He is also member of the Technical Program Committee (TPC) of the *IEEE International Symposium on Antennas and Propagation* since 2016 and has been member of the TPC of the *International Congress on Advanced Electromagnetic Materials in Microwaves and Optics-Metamaterials* during the years 2014-2016. He has also been serving as a Technical Reviewer of many high-level international journals related to electromagnetic field theory, metamaterials and plasmonics and he been selected as one of the Top Reviewers by the Editorial Board of the *IEEE Transactions on Antennas & Propagation* for several years.

His research interests include the design and the applications of microwave and optical artificially engineered materials and metasurfaces, the design of cloaking devices for scattering cancellation at microwave and optical frequencies with a particular emphasis on their applications to the antenna theory. His research activities resulted in more than 80 papers published in international journals and conference proceedings. Dr. Monti has been the recipient of some national and international awards and recognitions, including the *URSI Young Scientist Award* (2019), the *outstanding Associate Editor* of the *IEEE Transactions on Antennas and Propagation* (2019), the *Finmeccanica Group Innovation Award for young people* (2015) and the 2nd place at the student paper competition of the conference *Metamaterials'* (2012).

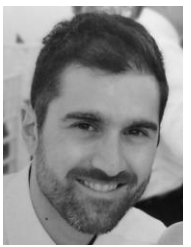

**Mirko Barbuto** (S'12–M'15–SM'19) was born in Rome, Italy, on April 26, 1986. He received the B.S., M.S. and Ph.D. degrees from ROMA TRE University, Rome, Italy, in 2008, 2010 and 2015, respectively. Since September 2013, he is with the "Niccolò Cusano" University, Rome, Italy, where he works as an Assistant Professor of electromagnetic field theory. His main research interests are in the framework of applied electromagnetics, with an emphasis on antennas and components at RF and microwaves, cloaking devices for radiating systems, metamaterials, electromagnetic structures loaded with non-linear or non-foster circuits, topological properties of vortex fields and smart antennas for GNSS technology.

He is a member of the Technical Program Committee of the *International Congress on Artificial Materials for Novel Wave Phenomena* (since 2017) and he serves as a Technical Reviewer of the major international conferences and journals related to electromagnetic field theory and metamaterials. He has been recipient of the *Outstanding Reviewers Award* assigned by the Editorial Board of the *IEEE Transactions on Antennas & Propagation* for four consecutive years (2015-2018) and he has received the same award by the Editorial Board of the *IEEE Antennas and Wireless Propagation Letters* for two consecutive years (2017-2018). In 2017 he has been selected as one of the *Best Reviewers* by the Editorial Board of *Radioengineering Journal*. Since 2015, he is the Proceeding Editor for the annual *International Congress on Engineered Material Platforms for Novel Wave Phenomena – Metamaterials*.

Dr. Barbuto is currently member of the *Italian Society on Electromagnetics* (SIEM), of the *National Inter-University Consortium for Telecommunications* (CNIT), of the *Virtual Institute for Artificial Electromagnetic Materials and Metamaterials* (METAMORPHOSE VI), and of the *Institute of Electrical and Electronics Engineers* (IEEE). Currently, he is the author of more than 50 papers in international journals and conference proceedings.

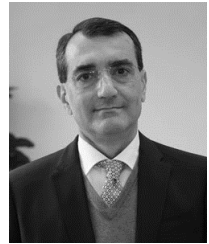

**Alessandro Toscano** (M'91-SM'11) (Capua, 1964) graduated in Electronic Engineering from Sapienza University of Rome in 1988 and he received his PhD in 1993. Since 2011, he has been Full Professor of Electromagnetic Fields at the Engineering Department of ROMA TRE University. He carries out an intense academic and scientific activity, both nationally and internationally.

From April 2013 to January 2018 he was a member of ROMA TRE University *Academic Senate*. From October 2016 to October 2018, he is a member of the National Commission which enables National Scientific Qualifications to Full and Associate Professors in the tender sector 09/F1 – Electromagnetic fields.

Since 23rd January 2018 he has been *Vice-Rector* for Innovation and Technology Transfer.

In addition to his commitment in organizing scientific events, he also carries out an intense editorial activity as a member of the review committees of major international journals and conferences in the field of applied electromagnetics. He has held numerous invited lectures at universities, public and private research institutions, national and international companies on the subject of artificial electromagnetic materials, metamaterials and their applications. He actively participated in founding the international association on metamaterials *Virtual Institute for Advanced Electromagnetic Materials –* METAMORPHOSE, VI. He coordinates and participates in several research projects and contracts funded by national and international public and private research institutions and industries.

Alessandro Toscano's scientific research has as ultimate objective the conceiving, designing and manufacturing of innovative electromagnetic components with a high technological content that show enhanced performance compared to those obtained with traditional technologies and that respond to the need for environment and human health protection. His research activities are focused on three fields: metamaterials and unconventional materials, in collaboration with Professor A. Alù's group at The University of Texas at Austin, USA, research and development of electromagnetic cloaking devices and their applications (First place winner of the *Leonardo Group Innovation Award* for the research project entitled: 'Metamaterials and electromagnetic invisibility') and the research and manufacturing of innovative antenna systems and miniaturized components (first place winner of the *Leonardo Group Innovation Award* for the research project entitled: "Use of metamaterials for miniaturization of components" – MiniMETRIS).

He is the author of more than one hundred publications in international journals indexed ISI or Scopus; of these on a worldwide scale, three are in the first 0.1 percentile, five in the first 1 percentile and twenty-five in the first 5 percentile in terms of number of quotations and journal quality.

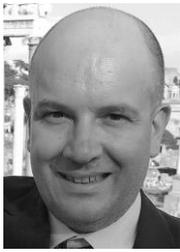

**Filiberto Bilotti** (S'97–M'02–SM'06–F'17) received the Laurea and Ph.D. degrees in electronic engineering from ROMA TRE University, Rome, Italy, in 1998 and 2002, respectively. Since 2002, he has been with the Faculty of Engineering (2002-2012) and, then, with the Department of Engineering (2013-now), at ROMA TRE University, where he serves as a Full Professor of electromagnetic field theory (2014-now) and the *Director of the Antennas and Metamaterials Research Laboratory* (2012-present)..

His main research contributions are in the analysis and design of microwave antennas and arrays, analytical modelling of artificial electromagnetic materials, metamaterials, and metasurfaces, including their applications at both microwave and optical frequencies. In the last ten years, Filiberto Bilotti's main research interests have been focused on the analysis and design of cloaking metasurfaces for antenna systems, on the modelling and applications of (space and) time-varying metasurfaces, on the topological-based design of antennas supporting structured field, on the modelling, design, and implementation of non-linear and reconfigurable metasurfaces, on the concept of meta-gratings and related applications in optics and at microwaves, on the modelling and applications of optical metasurfaces. The research activities developed in the last 20 years (1999-2019) has resulted in more than 500 papers in international journals, conference proceedings, book chapters, and 3 patents.

Prof. Bilotti has been serving the scientific community, by playing leading roles in the management of scientific societies, in the editorial board of international journals, and in the organization of conferences and courses.

In particular, he was a founding member of the *Virtual Institute for Artificial Electromagnetic Materials and Metamaterials* – METAMORPHOSE VI in 2007. He was elected as a member of the Board of Directors of the same society for two terms (2007-2013) and as the President for two terms (2013-2019). Currently, he serves the METAMORPHOSE VI as the Vice President and the Executive Director (2019-now).

Filiberto Bilotti served as an Associate Editor for the *IEEE Transactions on Antennas and Propagation* (2013-2017) and the journal *Metamaterials* (2007-2013) and as a member of the Editorial Board of the *International Journal on RF and Microwave Computer-Aided Engineering* (2009-2015), *Nature Scientific Reports* (2013-2016), and *EPJ Applied Metamaterials* (2013-now). He was also the Guest Editor of 5 special issues in international journals.

He hosted in 2007 the inaugural edition of the *International Congress on Advanced Electromagnetic Materials in Microwaves and Optics – Metamaterials Congress*, served as the Chair of the Steering Committee of the same conference for 8 editions (2008-2014, 2019), and was elected as the General Chair of the *Metamaterials Congress* for the period 2015-2018. Filiberto Bilotti was also the General Chair of the *Second International Workshop on Metamaterials-by-Design Theory, Methods, and Applications to Communications and Sensing* (2016) and has been serving as the chair or a member of the technical program, steering, and organizing committee of the main national and international conferences in the field of applied electromagnetics.

Prof. Bilotti was the recipient of a number of awards and recognitions, including the elevation to the *IEEE Fellow* grade for contributions to metamaterials for electromagnetic and antenna applications (2017), *outstanding Associate Editor of the IEEE Transactions on Antennas and Propagation* (2016), *NATO SET Panel Excellence Award* (2016), *Finmeccanica Group Innovation Prize* (2014), *Finmeccanica Corporate Innovation Prize* (2014), IET Best Poster Paper Award (Metamaterials 2013 and Metamaterials 2011), Raj Mittra Travel Grant Senior Researcher Award (2007).